\documentclass[onecolumn,a4paper,10pt,showpacs,amsmath,amssymb]{revtex4}
\pdfoutput=1

\usepackage{ifpdf}

 \ifpdf
    \usepackage[pdftex]{graphicx}
 \else
    \usepackage[dvips]{graphicx}
 \fi

\ifpdf
       \DeclareGraphicsExtensions{.pdf,.png}
 \else
       \DeclareGraphicsExtensions{.eps}
\fi

\usepackage[colorlinks]{hyperref}

\begin{document}

\title{Nonlinear Electron Oscillations in a Viscous and Resistive Plasma}

\author{A. A. Skorupski}
\email[]{askor@fuw.edu.pl}
\affiliation{Department of Theoretical Physics, Soltan Institute for Nuclear
Studies, Ho\.za 69, 00--681 Warsaw, Poland}

\author{E. Infeld}
\email[]{einfeld@fuw.edu.pl}
\affiliation{Department of Theoretical Physics, Soltan Institute for Nuclear
Studies, Ho\.za 69, 00--681 Warsaw, Poland}

\date{\today}

\begin{abstract}
New non-linear, spatially periodic, long wavelength electrostatic modes of an electron
fluid oscillating against a motionless ion fluid (Langmuir waves) are given, with
viscous and resistive effects included. The cold plasma approximation is adopted, which
requires the wavelength to be sufficiently large. The pertinent requirement valid for
large amplitude waves is determined. The general non-linear solution
of the continuity and momentum transfer equations for the electron fluid along with
Poisson's equation is obtained in simple parametric form. It is shown that in all
typical hydrogen plasmas, the influence of plasma resistivity on the modes in question
is negligible.
Within the limitations of the solution found, the non-linear
time evolution of \textit{any} (periodic) initial electron number density profile
$n_e(x, t=0)$ can be determined (examples).
For the modes in question, an idealized model of a strictly cold and collisionless plasma
is shown to be applicable to any real plasma, provided that the wavelength
$\lambda \gg \lambda_{\text{min}}(n_0,T_e)$, where $n_0 = \text{const}$
and $T_e$ are the equilibrium values of the electron number density and electron
temperature. Within this idealized
model, the minimum of the initial electron density $n_e(x_{\text{min}}, t=0)$ must be
larger than half its equilibrium value, $n_0/2$. Otherwise, the corresponding
maximum $n_e(x_{\text{max}},t=\tau_p/2)$, obtained after half a period of the plasma
oscillation blows up. Relaxation of this restriction on $n_e(x, t=0)$ as one
decreases $\lambda$, due to the increase of the electron viscosity effects, is examined
in detail. Strong plasma viscosity is shown to change considerably the density profile
during the time evolution, e.g., by splitting the largest maximum in two.
\end{abstract}

\maketitle

\section{Introduction}

In this paper, which is an extension and generalisation of an earlier Letter \cite{prl09},
we present new features of an important mode in a plasma, i.e., electron plasma
waves, also called Langmuir oscillations or space charge waves. The non-linear
(fluid) features of these waves will be examined in full detail, along with
approximate inclusion of the dissipative effects due to electron viscosity and plasma
resistivity.

Electron plasma waves were first examined, both experimentally and theoretically, in
1929 by Tonks and Langmuir. A simple theory was given \cite{TL}, treating the plasma
electrons as non-interacting particles oscillating against the motionless ions.
% under the influence of the electric field due to charge separation.
This simple picture, neglecting random thermal motions, led to the equation of
motion for each electron to be that for a harmonic oscillator with
angular frequency
\begin{equation}
\omega^2 = \omega_{pe}^2 = 4\pi n_0 e^2/m_e,\label{omegape}
\end{equation}
where $e$ and $m_e$ are the electron charge and mass, and $n_0 = \text{const}$
is the equilibrium value of the electron number density. A similar equation was obtained
for the electric field $\mathbf{E}$ under the assumption that $\nabla\times
\mathbf{E} = 0$, i.e., that there is no magnetic field associated with the wave.
This means that the electric current due to electron oscillations is fully
compensated by the displacement current $\dot{\mathbf{E}}/(4\pi)$. This fact and 
formula (\ref{omegape}) defining the frequency of electron oscillations are the
most essential features of Langmuir oscillations in the ``cold plasma approximation'',
neglecting thermal motions.

The influence of random thermal motions on small oscillations of electrons can be
examined within the kinetic theory based on the linearised Vlasov equation. This was
done correctly for the first time in 1946 by Landau \cite{landau} who arrived at the
dispersion relation in the form
\begin{equation}
\omega^2 = \omega_{pe}^2 + k^2 (3T_e/m_e) \equiv \omega_{pe}^2 [1 + 3 (k\lambda_D)^2],
\label{disprBG}
\end{equation}
valid in the long wavelength limit, $(k\lambda_D)^2 \ll 1$, where $k = 2\pi/\lambda$ is
the wavenumber, $\lambda$ is the wavelength, $\lambda_D = v_{\text{th}}/\omega_{pe}$ is
the Debye length, $v_{\text{th}} = \sqrt{T_e/m_e}$ is the electron thermal speed, and
$T_e$ is the equilibrium electron temperature in energy units.
% (i.e., actually stands for $k_B T_e$ where $k_B$ is the Boltzmann constant).
In this long wavelength limit,
the phase velocity of the wave, $v_{\text{ph}} = \omega/k \approx \omega_{pe}/k$
is much larger than $v_{\text{th}}$. Landau's analysis also indicated that
the electron plasma waves in this limit undergo a small
collisionless damping with decrement given by
\begin{equation}
\omega_{pe} \sqrt{\frac{\pi}{8}} \frac{1}{(k\lambda_D)^3} \exp\biggl[ -
\frac{1}{2(k\lambda_D)^2} \biggr].
\label{Ldamp}
\end{equation}
These results, along with the contour integrals used in the derivations, turned out to be
very important for further progress in the kinetic theory of various waves in plasmas,
see e.g., \cite{krall}. The decrement given by (\ref{Ldamp}) is nowadays
referred to as Landau damping. The dispersion relation (\ref{disprBG})
is commonly associated with the names of Bohm and Gross who derived it independently
three years later \cite{BG}. However, their derivation, not based on either Vlasov
theory or any other systematic method, had no chance of generalisations or finding damping.

The physics behind Landau damping is nowadays well understood. It appears for any
particle distribution function which decreases as one increases the particle velocity
in some vicinity of the phase velocity of the wave. In that case there are more
particles that are a bit slower than the wave and are accelerated on account of the wave
energy than those decelerated. Large Landau damping is obtained if the phase velocity of
the wave is close to the particle thermal speed where the slope of the distribution
function is large.

The simplest model that can be used to examine waves in a plasma is the fluid model in
which electrons and each kind of ions are treated as fluids. In the small amplitude limit
which allows for linearisation of the macroscopic equations for $n_e - n_0$,
$\mathbf{v}_e$, etc., one is dealing with linear and homogeneous partial differential
equations for these parameters and the electric and magnetic fields,  see e.g.,
\cite{linwav}. Very fine effects, like Landau damping, are
beyond the scope of the fluid description. However, it is much easier to include non-linear
and dissipative effects, the presence of external fields in equilibrium or equilibrium
inhomogeneities.

If the equilibrium parameters are space independent, the partial differential equations
in question have constant coefficients and are satisfied by plane waves, where
$n_e - n_0$, $\mathbf{v}_e$, $\mathbf{E}$ etc., are proportional to
$\exp[i(\mathbf{k}\cdot\mathbf{r} - \omega t)]$. In the case of Langmuir oscillations,
the waves are longitudinal ($\mathbf{E}$ and $\mathbf{v}_e$ have the direction of the wave
vector $\mathbf{k}$, due to $\nabla\times \mathbf{E} = 0$, i.e., $\mathbf{k} \times
\mathbf{E} = 0$)). One can always choose the $x$ axis of the coordinate system to have
the direction of $\mathbf{k}$. With this
choice, $\mathbf{k}\cdot\mathbf{r} = k x$ and the vectors $\mathbf{v}_e$ and $\mathbf{E}$
have $x$ components only. The real part of such a solution is one-dimensional, periodic
in $x$ with wavelength $\lambda = 2\pi/k$. Its profile is purely cosinusoidal. More
complicated periodic profiles with the same $\lambda$ can be obtained by
adding higher harmonics, with $k$ replaced by $n k$, $n = 1, 2, \dots$ In the linearised
theory, they
will also be solutions, oscillating  with $\omega = \omega_{pe}$. The aim of this paper
is to describe the behaviour of these 1D solutions as one increases
the amplitudes so that the non-linear terms cannot be neglected, and also with dissipative
effects included.

For a purely cosinusoidal initial profile of $n_e - n_0$ and no dissipation, this problem was
solved formally  by Dawson in 1959 by introducing a Lagrangian coordinate and in more detail
by Davidson and Schram in 1968 \cite{david}. The solution had a somewhat unexpected initial
amplitude limitation. This will follow from our results as a very special case.

In the linearised theory one can superpose plane plasma waves propagating in various
directions. Such higher dimensional solutions are beyond
the scope of our fully non-linear 1D analysis. They are not so important as the 1D Langmuir
waves that are importantly recently revisited in connection with laser-driven plasma-based
electron accelerators. Such accelerators are capable of supporting fields even in excess of
100 GV/m \cite{plasacc}.

While in the kinetic theory of Langmuir waves as a rule one assumes that ions are
motionless, in the fluid model one can easily include motions of all plasma
species, to check this assumption quantitatively. To close the set of fluid
equations involving the continuity and momentum transfer equations for all
species, one has to postulate a pressure--density relation. If the polytropic
one is assumed ($p_{\alpha}/n_{\alpha}^{\gamma_{\alpha}} = \text{const}$), the
relevant equations for the longitudinal waves propagating along the $x$ axis are
as follows \cite{linwav} ($\alpha = e$ for electrons and $1,2,\dots,N_i$ for ions).
\\[1ex]
%\noindent
Dispersion relation:
\begin{equation}
1 - \sum_{\alpha} \frac{(\omega_{p \alpha}/\omega)^2}{1 - \gamma_{\alpha}
(v_{\text{th} \, \alpha}/v_{\text{ph}})^2} = 0, \quad
\omega_{p \alpha}^2 = \frac{4\pi n_{\alpha 0} \, q_{\alpha}^2}{m_{\alpha}}.
\label{flmdl1}
\end{equation}
Complex amplitudes of the macroscopic velocity $v_{\alpha}$ versus electric field
$E$:\\
\begin{equation}
v_{\alpha} = i \frac{q_{\alpha}}{\omega m_{\alpha}}\,\frac{E}{1 - \gamma_{\alpha}
(v_{\text{th} \, \alpha}/v_{\text{ph}})^2},
\label{flmdl2}
\end{equation}
where $q_e = -e$, $q_i = Z_i e$, $v_{\text{th} \, \alpha} =
\sqrt{T_{\alpha}/m_{\alpha}}$, $v_{\text{ph}} = \omega/k$, $n_{\alpha 0}$ and
$T_{\alpha}$ are the equilibrium values of the number density and temperature in
energy units for the $\alpha$ species ($n_{e 0} \equiv n_0$).
The thermal corrections (subtracted from 1 in the denominators) must be small as
compared to unity, to avoid strong Landau damping. Neglecting them we obtain
\begin{equation}
\omega^2 = \omega_p^2 \equiv \sum_{\alpha} \omega_{p \alpha}^2 =
\omega_{pe}^2 \biggl[1 + \sum_{i = 1}^{N_i}
\text{O}(m_e/m_i) \biggr],
\label{flmdl3}
\end{equation}
\begin{equation}
v_i = - Z_i \, \frac{m_e}{m_i} \,  v_e.
\label{flmdl4}
\end{equation}
Eqs.~(\ref{flmdl3}) and (\ref{flmdl4}) justify the assumption of motionless ions
($m_i \to \infty$). Under this assumption, Eq.~(\ref{flmdl1}) simplifies to
\begin{equation}
\omega^2 = \omega_{pe}^2 + k^2 (\gamma_e T_e/m_e).
\label{disprfm}
\end{equation}
This dispersion relation following from the fluid model will coincide with that
from the kinetic theory, Eq.~(\ref{disprBG}), if we choose $\gamma_e = 3$. This
value is thus appropriate for the fluid description of Langmuir waves, which can
be interpreted as a one-dimensional adiabatic compression of the electron gas,
associated with these waves.

\section{Basic equations}

For $x$ dependent electron plasma waves and a motionless ion fluid, the
model equations describing the electron fluid are: the continuity and
momentum transfer equations along with the adiabatic pressure--density relation
with $\gamma_e = 3$, and  Poisson's equation, see e.g., \cite{brag} (Gaussian units),
\begin{eqnarray} 
\frac{\partial n_e}{\partial t} + \frac{\partial}{\partial x} \bigl(
n_e v_e \bigr) &=& 0,\label{cont}\\
m_e n_e \Bigl( \frac{\partial v_e}{\partial t} + v_e \frac{\partial v_e}{\partial
x} \Bigr) &=& -\frac{\partial p_e}{\partial x} - e n_e E +
\frac{\partial}{\partial x} \biggl( \frac{4}{3} \nu_e
\frac{\partial v_e}{\partial x} \biggr) - \eta \, e^2 n_0 n_e v_e,\label{mom}\\
\frac{p_e}{n_e^3} &=& \frac{p_0}{n_0^3} = \frac{T_e}{n_0^2}\\
\frac{\partial E}{\partial x} &=& 4 \pi e ( n_0 - n_e ),\label{pois}
\end{eqnarray}
where the equilibrium electron temperature $T_e$ is in energy units, $\nu_e$ is the
electron viscosity coefficient, $\eta$ is the plasma resistivity, and an ideal gas
equation of state for the electron gas in equilibrium is assumed, $p_0 = n_0 T_e$.

We assume that the first term on the right hand side in Eq.~(\ref{mom}) is
negligible as compared to the second one, i.e., that the wave is driven by
the electric field rather than by electron pressure (cold plasma approximation).
For small amplitudes that
allow for linearisation, that will be the case if the thermal correction in the
dispersion relation (\ref{disprBG}) is negligible, i.e., 
\begin{subequations}
\label{valc1}
\begin{equation}
3 \, (k \lambda_D)^2 \equiv \Bigl(\frac{3\pi}{e^2}\Bigr)
\frac{T_e}{n_0 \lambda^2}  \ll 1, \quad \text{where}
\end{equation}

\begin{equation}
k\lambda_D = 2\pi \, \frac{\lambda_D}{\lambda} =
\frac{k v_{\text{th}}}{\omega_{pe}} \approx
\frac{k v_{\text{th}}}{\omega} \equiv \frac{v_{\text{th}}}{v_{\text{ph}}}
\equiv \frac{\tau_p v_{\text{th}}}{\lambda} \, ,
\end{equation}
\end{subequations}
and $\tau_p = 2\pi/\omega$ is the period of plasma oscillations. This
condition is fulfilled in any plasma, provided that the wavelength $\lambda$
is sufficiently large as compared to the Debye length. In that case, $\lambda$
is also large as compared to the distance travelled by an electron moving with
its thermal speed during one period of plasma oscillation, and phase velocity
of the wave is large as compared to the electron thermal speed. The non-linear
counterpart of condition (\ref{valc1}) will be derived in Sec.~\ref{applc}.

Analytical formulas for the electron viscosity coefficient $\nu_e$ and
resistivity $\eta$ involve several approximations, see Sec.~\ref{applc}.
Using the latest values of the physical constants \cite{physpar}, the relevant
formulas given in the first reference of \cite{brag} take the form (two
component plasma of electrons and one kind of $Z=1$ ions, i.e., protons,
deuterons, or tritons):
\begin{subequations}
\label{trco}
\begin{equation}
\nu_e = 0.73 \frac{3 \sqrt{m_e} T_e^{5/2}}{4 \sqrt{2\pi} \lambda_C e^4} =
4.01318\!\times\!10^{-8} \frac{(T_e[\text{eV}])^{5/2}}{\lambda_C/10},\label{nue}
\end{equation}
\begin{equation}
\eta = 0.51 \frac{4\sqrt{2\pi m_e} \lambda_C e^2}{3 T_e^{3/2}} = 5.86059%
\!\times\!10^{-14} \frac{\lambda_C/10}{(T_e[\text{eV}])^{3/2}},\label{eta}
\end{equation}
\end{subequations}
where $\lambda_C$ is a slowly varying Coulomb logarithm which will be treated as
a constant:
\begin{equation}
\lambda_C = \begin{cases}
23.4 - \ln(n_0^{1/2}\,T_e^{-3/2})& \text{if $T_e \leq 50$ eV},\\
25.356 - \ln(n_0^{1/2}\,T_e^{-1})& \text{if $T_e \geq 50$ eV}.
\end{cases}\label{coul}
\end{equation}
(Typically $\lambda_C \approx$ 10--20). 

Neglecting the electron pressure term,
we can write Eq.~(\ref{mom}) in the form
\begin{equation}
\frac{\partial v_e}{\partial t} + v_e \frac{\partial v_e}{\partial x} =
-\frac{e}{m_e}\,E + \frac{\nu}{m_e n_e} \frac{\partial^2 v_e }{\partial x^2} -
\eta \, \frac{e^2 n_0}{m_e} \, v_e.\label{veq} 
\end{equation} 
where $\nu = (4/3)\nu_e$.

\section{New variables}

The nonlinearity on the left hand side of Eq.~(\ref{veq}) can be eliminated by
introducing Lagrangian coordinates: $x_0(x,t)$, the initial position (at $t=0$)
of an electron fluid element which at time $t$ was at $x$, and time in the
electron fluid rest frame, $\tau (= t)$. The basic transformation equations
between Eulerian and Lagrangian coordinates are (see \cite{infbook} for more
detail)
\begin{equation}\label{trans}
\frac{\partial}{\partial \tau} = \frac{\partial}{\partial t} + v_e
\frac{\partial}{\partial x}, \quad x = x_0 +
\int_0^{\tau}v_e(x_0,\tau')\,d\tau'.
\end{equation}
The continuity equation (\ref{cont}) in the electron fluid rest frame is now simply
\begin{equation}\label{cont1}
n_e \frac{\partial x}{\partial x_0} = n_{e0}(x_0) \equiv n_e(x_0, t=0).
\end{equation}
This indicates that the nonlinear operator $n_e^{-1}\partial/\partial x$ ($=
n_{e0}^{-1}\partial/\partial x_0$) in Eq.~(\ref{veq}) gets a bit simpler in the
Lagrangian coordinates. However, to make it linear, a new space variable $s$ has
to be introduced satisfying:
\begin{equation}\label{ss}
\frac{1}{n_e(x,t)}\frac{\partial}{\partial x} =
\frac{1}{n_{e0}(x_0)}\frac{\partial}{\partial x_0} =
\frac{1}{n_0}\frac{\partial}{\partial s}.
\end{equation}
The second equality here implies the definition of the auxiliary variable $s$:
\begin{equation}\label{svar} 
s = n_0^{-1}\int n_{e0}(x_0)\,dx_0,
\end{equation}
which turns out to be proportional to the integral of the initial electron density
profile.

Using Eqs.~(\ref{trans}) and (\ref{ss}), Eq.~(\ref{veq}) takes the form
\begin{equation}
\frac{\partial v_e}{\partial \tau} + \frac{e}{m_e}\,E - \frac{\nu}{m_e n_0}
\frac{\partial}{\partial s}\Bigl[\Bigl(\frac{n_e}{n_0}\Bigr)
\frac{\partial v_e}{\partial s}\Bigr] + \eta \, \frac{e^2 n_0}{m_e} \, v_e= 0.
\label{veql} 
\end{equation} 
An important point is that $E$ can be linearly expressed in terms
of $v_e$. Indeed, using  Eqs.~(\ref{pois}) and (\ref{cont}) it follows that
\begin{equation}
\label{dEdt}
\frac{\partial E}{\partial t} = 4\pi e n_e v_e ,
\end{equation}
which expresses the fact that the electric current, $-e n_e v_e$, is
compensated by the displacement current $(4\pi)^{-1}\partial E/\partial t$.
Adding this equation to $v_e$ times Eq.~(\ref{pois}) we end up with

\begin{equation}
\label{dEdta} 
\frac{\partial E}{\partial\tau} = 4\pi e n_0 v_e ,
\end{equation}
where the right hand side is linear in $v_e$ as promised. Therefore, if we
linearise the viscous term in Eq. (\ref{veql}) by replacing $n_e$ by $n_0$,
differentiate this equation $\partial/\partial\tau$ and use (\ref{dEdta}),
we obtain 
\begin{equation} 
\frac{\partial^2 v_e}{\partial \tau^2} - \frac{\nu}{m_e n_0} 
\frac{\partial^2}{\partial s^2}\frac{\partial v_e}{\partial \tau}
+ \eta \, \frac{\omega_{pe}^2}{4\pi} \frac{\partial v_e}{\partial\tau}
+ \omega_{pe}^2 v_e = 0,\label{veql1}
\end{equation} 
where $\omega_{pe}^2$ is defined by Eq.~(\ref{omegape}).
Linearisation of the viscous term is a simplification justified by the fact that
this term only corrects the motion, whereas the driving electric force and the
convective non-linearities in the fluid equations are taken into account exactly.
Furthermore, the formula for the viscosity coefficient, Eq.~(\ref{nue}), is
approximate, up to a factor of two or three.

Eq.~(\ref{veql1}) is a \textit{linear} partial differential equation (PDE) for
$v_e(s,\tau)$ with \textit{constant} coefficients. Solutions of such PDEs are any
superpositions of the normal modes $\exp[i(ks - \Omega\tau)]$, for which
Eq.~(\ref{veql1}) leads to the dispersion relation

\begin{subequations}
\label{disp} 
\begin{eqnarray}
\Omega &=& \pm \omega(k) - i \alpha(k),\\
\alpha(k) &=& \frac{\nu k^2}{2 m_e n_0} + \frac{\eta \, e^2 n_0}{2 m_e},\\
\omega(k) &=& \sqrt{\omega_{pe}^2 - \alpha^2(k)}.
\end{eqnarray}
\end{subequations}
Assuming that $k$ is real and superposing the normal modes corresponding to
the plus and minus signs in $\Omega$ given by Eq.~(\ref{disp}) we obtain four
real solutions:
\begin{equation} 
v_e = e^{-\alpha(k)\tau} f\bigl(\omega(k) \tau\bigr) \, g(ks),\label{vsol}
\end{equation} 
where $f,g = \sin \: \text{or} \: \cos$.

For each $\tau$, $v_e$ given by Eq.~(\ref{vsol}) is a periodic function of $s$
with wavelength $\lambda=2\pi/k$. By adding higher harmonics, obtained from
Eq.~(\ref{vsol}) on replacing $k \to nk$, $n = 1,2,\ldots$, and multiplying by
an amplitude, any solution periodic in $s$ with wavelength $\lambda$ can be
obtained. Replacing $a\sin(\omega\tau) + b\cos(\omega\tau) \equiv
A\cos(\omega\tau + \varphi)$, the solution in question can be written as
\begin{eqnarray} 
v_e &=& \sum_{n=1}^{\infty}\: e^{ -\alpha(nk) \tau} \:\Bigl\{ A_{1n}
\cos\Bigl[\omega(nk)\tau + \varphi_{1n} \Bigr]\sin(nks) + A_{2n}
\cos\Bigl[\omega(nk)\tau + \varphi_{2n} \Bigr]\cos(nks)\Bigr\},\nonumber\\
&&\label{vsum}
\end{eqnarray} 
where $A_{1n}$, $\varphi_{1n}$,  $ A_{2n}$, and $\varphi_{2n}$ are arbitrary
constants.

\section{General solution in parametric form}

Our equations and final results take a simple and universal form if we
introduce dimensionless quantities:
\begin{subequations}
\label{dimlpar}
\begin{eqnarray}
\bar{x} &=& kx, \quad \bar{s} = ks, \quad \bar{t} = \omega_{pe} t, \quad
\bar{\tau} = \omega_{pe} \tau,\\
\bar{\omega}_n &=& \frac{\omega(nk)}{\omega_{pe}} = \sqrt{1 - \bar{\alpha}_n^2},
\quad  \bar{\alpha}_n = n^2 \bar{\nu} + \bar{\eta},\label{alphab}\\
\bar{\eta} &=& \!\frac{\eta \, n_0 e^2}{2 m_e\omega_{pe}} = 1.31642\!\times\! 10^{-10}
\frac{\lambda_C}{10} \frac{n_0^{1/2}}{(T_e[\text{eV}])^{3/2}},\label{etab}\\
\bar{\nu} &=& \frac{2\nu_e k^2}{3 m_e n_0 \omega_{pe}} = 2.70376\!\times\! 10^6
\frac{T_e[\text{eV}]}{\bar{\eta} n_0 \lambda^2},\label{nub}\\
\bar{v}_e &=& \frac{v_e}{v_{\text{ph}}}, \quad v_{\text{ph}} =
\frac{\omega_p}{k},\\
\bar{A}_{1n} &=&
\frac{A_{1n}}{v_{\text{ph}}}, \quad \bar{A}_{2n} =
\frac{A_{2n}}{v_{\text{ph}}},\\
\bar{n}_e &=& \frac{n_e}{n_0}, \quad \bar{E} = \frac{E e k}{m_e\omega_{pe}^2}.
\end{eqnarray}
\end{subequations}
Thus, Eqs.~(\ref{pois}), (\ref{ss}) and (\ref{dEdta}) are now:
\begin{equation}
\frac{\partial\bar{E}}{\partial\bar{x}} = 1 - \bar{n}_e, \quad
\frac{\partial\bar{s}}{\partial\bar{x}} = \bar{n}_e, \quad
\frac{\partial\bar{E}}{\partial\bar{\tau}} = \bar{v}_e.\label{dimleqs}
\end{equation}
Integrating the first two over $d \bar{x}$ and the last one over
$d \bar{\tau}$ and using (\ref{vsum}), we end up with equations
which define all relevant quantities: $\bar{x}$, $\bar{E}$,
$\bar{n}_e$ and $\bar{v}_e$ in terms of $\bar{s}$ and $\bar{\tau}$ ($=\bar{t}$).
Dropping bars for simplicity, the final results for the dimensionless quantities
become:
\begin{subequations}
\label{dimles}
\begin{eqnarray}
x(s,t) &=& s + E ,\label{barx}\\
E(s,t) &=& - \sum_{n=1}^{\infty} \: e^{ -\alpha_n t} \:\Bigl[ A_{1n}g_{1n}(t)\sin(ns) 
+ A_{2n}g_{2n}(t)\cos(ns)\Bigr],\label{barE}\\
n_e^{-1}(s,t) &=& 1 - \sum_{n=1}^{\infty} \: e^{ -\alpha_{n} t} \: n
\Bigl[ A_{1n}g_{1n}(t)\cos(ns) - A_{2n}g_{2n}(t)\sin(ns)\Bigr],\label{barne}\\
v_e(s,t) &=& \sum_{n=1}^{\infty} \: e^{ -\alpha_n t} \:
\Bigl[ A_{1n}f_{1n}(t)\sin(ns) 
+ A_{2n}f_{2n}(t)\cos(ns)\Bigr],\label{vsumd}\\
f_{jn}(t) &=& \cos(\omega_n t + \varphi_{jn}), \quad j = 1,2,\nonumber\\
g_{jn}(t) &=& \alpha_n f_{jn}(t) - \omega_n \sin(\omega_n t +
\varphi_{jn}) \equiv \cos(\omega_n t + \varphi_{jn} + \varphi_{0n}),\label{fgjn}\\
\varphi_{0n} &=& \arctan(\omega_n/\alpha_n).\label{phi0n}
\end{eqnarray}
\end{subequations} 
The identity in Eq.~(\ref{fgjn}) is due to the fact that $\alpha_n^2 + \omega_n^2 =
1$, and therefore one can always find such $\varphi_{0n}$ that $\cos(\varphi_{0n}) =
\alpha_n$, and $\sin(\varphi_{0n}) = \omega_n$.

The independent parameters $A_{1n}$, $\varphi_{1n}$, $ A_{2n}$, and $\varphi_{2n}$
must be chosen so as to ensure reality of all dependent parameters defined by
 Eqs.~(\ref{dimles}), $x$, $E$, $n_e$, and $v_e$.

For modes with $n$ not too large, such that $\alpha_n \equiv n^2 \nu + \eta < 1$,
$\omega_n$ and $\varphi_{0n}$ are real and positive, and $f_{jn}(t)$ and
$g_{jn}(t)$ are purely oscillating cosine functions, with period $2\pi/\omega_n$ and
arguments shifted by $\varphi_{0n}$.

For higher modes, either $\omega_n = \varphi_{0n} = 0$, if $\alpha_n = 1$, leading
to $f_{jn}(t) = g_{jn}(t) = \cos(\varphi_{jn}) = \text{const}$, or $\omega_n$ and
$\varphi_{0n}$ become pure imaginary ($= i\lvert\omega_n\rvert$ and
$i\lvert\varphi_{0n}\rvert$), if $\alpha_n > 1$. Only the latter case requires
further attention, whereas in the remaining cases ($\alpha_n \leq 1$) we can choose
$A_{jn}$ to be real, $j=1,2$.

Thus if $\alpha_n > 1$, real values of the phase shifts $\varphi_{jn}$ in general
lead to $f_{jn}(t)$ and $g_{jn}(t)$ being complex and having time dependent phases.
Therefore in that case one cannot produce real results by an appropriate choice of
complex coefficients $A_{1n}$ and $A_{2n}$. The only exceptions are $\varphi_{jn}
= 0 \text{ or} -\varphi_{0n}$ and $\varphi_{jn} = (0 \text{ or} -\varphi_{0n})
\mp\pi/2$.

If we choose $\varphi_{jn} = 0$, the resulting $f_{jn}(t) = \cosh(\lvert\omega_n\rvert
t)$ and $g_{jn}(t) = \cosh(\lvert\omega_n\rvert t + \lvert\varphi_{0n}\rvert)$ will be
real. A similar situation will arise if we choose $\varphi_{jn} = -\varphi_{0n}$, for
which $f_{jn}(t) = \cosh(\lvert\omega_n\rvert t - \lvert\varphi_{0n}\rvert)$,
$g_{jn}(t) = \cosh(\lvert\omega_n\rvert t)$. In all these cases we can choose $A_{jn}$
to be real, $j=1,2$.

If we choose $\varphi_{jn} = \mp \pi/2$ (and also for $\varphi_{jn} =
 -\varphi_{0n} \mp \pi/2$), it is convenient to replace the amplitudes
$A_{jn}$ by $A_{jn}/\omega_n$. With this choice Eq.~(\ref{dimles}) will hold if new
definitions of the functions $f_{jn}(t)$ and $g_{jn}(t)$ are adopted, i.e.,
\begin{subequations}
\label{spc1}
\begin{eqnarray}
f_{jn}(t)&=& \pm \frac{\sin(\omega_n t)}{\omega_n},\\
g_{jn}(t)&=& \pm \Bigl[ \alpha_n \frac{\sin(\omega_n t)}{\omega_n} +
\cos(\omega_n t)\Bigr],
\end{eqnarray}
\end{subequations} 
or
\begin{subequations}
\label{spc2}
\begin{eqnarray}
f_{jn}(t)&=& \pm \Bigl[ \alpha_n \frac{\sin(\omega_n t)}{\omega_n} -
\cos(\omega_n t),\Bigr]\\
g_{jn}(t)&=& \pm \frac{\sin(\omega_n t)}{\omega_n},
\end{eqnarray}
\end{subequations} 
where $\frac{\sin(\omega_n t)}{\omega_n}$ must be replaced by $t$ if $\omega_n = 0$.
It can be seen that the right hand sides of Eqs.~(\ref{spc1}) and (\ref{spc2}) are well
defined and real for any $n$ (if $\alpha_n > 1$, $\cos(\omega_n t) \equiv
\cosh(\lvert\omega_n\rvert t)$ and $\sin(\omega_n t)/\omega_n \equiv
\sinh(\lvert\omega_n\rvert t)/\lvert\omega_n\rvert$).
Their values at $t=0$ are either zero or $\pm 1$.
Thus, if Eq.~(\ref{spc1}) holds for both $j=1$ and $j=2$, this corresponds
to the unperturbed value of the initial electron velocity ($v_e(x,t=0)\equiv 0$), and
maximally perturbed initial values of the electric field $E$ and $n_e$.
If in that case ($\varphi_{jn} = - \pi/2$) we choose $A_{2n} \equiv 0$ and $A_{1n} =
A_n$, we will reproduce our earlier results given in \cite{prl09}. Similarly,
Eq.~(\ref{spc2}) corresponds to the unperturbed initial values
of the electric field and the electron density ($E(x,t=0) \equiv 0$ and $n_e(x,t=0) = 1$),
and maximally perturbed initial value of the electron velocity.

Real values of $f_{jn}(t)$ and $g_{jn}(t)$ for $\alpha_n > 1$ will also be obtained
if we choose $\varphi_{jn}$ to be pure imaginary.

Eq.~(\ref{barne}) is only meaningful if the sum is smaller than unity, which
imposes a limitation on the amplitudes $A_{1n}$ and $A_{2n}$.

If $\varphi_{1n} = \varphi_{2n}$, the mode in question is a
product of a function of $t$ and a function of $s$, i.e., represents a standing wave.

Note that all modes, both with periodic $f_{jn}(t)$ and $g_{jn}(t)$ (for 
$\alpha_n < 1$) and aperiodic ones (for $\alpha_n \geq 1$) are damped exponentially
as $t \to \infty$ (with decrement $\alpha_n$ if $\alpha_n \leq 1$ or $\alpha_n -
\sqrt{\alpha_n^2 - 1} \equiv [\sqrt{\alpha_n^2 - 1} + \alpha_n]^{-1}$ if
$\alpha_n \geq 1$).

\section{Solution given in terms of physical variables}

\begin{figure}[t!]
\begin{center}

\ifpdf
    \includegraphics[type=png,ext=.png,read=.png,scale=0.6]{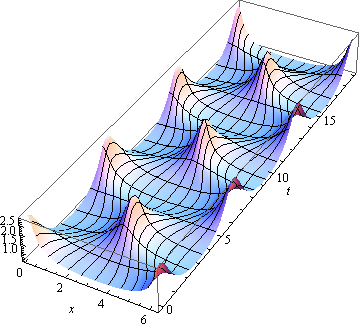}
\else
    \includegraphics[type=eps,ext=.eps,read=.eps,scale=0.9]{A1}
\fi 
%\centerline{\includegraphics[scale=.9]{A1.eps}}
\caption{(color online). Plot of $n_e(x,t)$ when only
 $A_{11} = 0.65$ is nonzero, and $\nu = 0$. This describes all situations with
 $\nu \ll 1$. The corresponding $\lambda$, for given $n_0$ and $T_e$, can be found
 from Eqs.~(\ref{etab}), (\ref{nub}) or Table~\ref{tableta}.\label{A1neq0}}
\end{center}
\end{figure}

\begin{figure}[b!]
\begin{center}

\ifpdf
    \includegraphics[type=png,ext=.png,read=.png,scale=0.6]{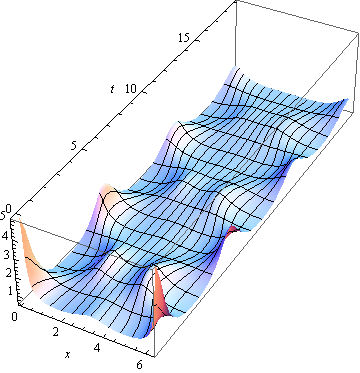}
\else
    \includegraphics[type=eps,ext=.eps,read=.eps,scale=0.9]{A1A2}
\fi 
%\centerline{\includegraphics[scale=.36]{A1A2.eps}}
\caption{(color online). Plot of $n_e(x,t)$ when only
$A_{11} = 0.4$ and $A_{12} = 0.2$ are nonzero, and $\nu = 0.02$.
($\sqrt{\nu^{-1}} = 7.07$, see Table~\ref{tableta}.) The viscous damping
is evident after a single period.\label{A1A2neq0}
}
\end{center}
\end{figure}

One can eliminate the parameter $s$ from Eqs.~(\ref{dimles}) thereby making $x$
and $t$ the independent variables. This parameter has to be determined in terms
of $x$ and $t$ from Eq.~(\ref{barx}) and used in  the remaining
Eqs.~(\ref{dimles}). While numerically this is a simple task, analytical
formulas are complicated, see Eqs.~(\ref{nefx})--(\ref{neve}). At the same
time, the parametric form (\ref{dimles}), which involves
simple elementary functions, can also be used to plot $E$, $n_e$ and
$v_e$ as functions of the physical variables $x$ and $t$ (see
Figs.~\ref{A1neq0}--\ref{B1B2lgv} and \ref{stfpp}--\ref{sawln} obtained
by using ``ParametricPlot3D'' of Mathematica 7).

The electron density $n_e(x,t)$ is a periodic function of $x$ with
wavelength $2\pi$. It can be expanded in a Fourier series in $x$ with time
dependent Fourier coefficients:
\begin{equation}
\label{nefx}
n_e(x,t) = 1 + \sum_{m=1}^{\infty} \Bigl[B_{1m}(t) \cos(m x) + B_{2m}(t) \sin(m x)
\Bigr],
\end{equation}
where ($n_e \, dx = ds$)
\begin{equation}
B_{1m}(t) \equiv \frac{1}{\pi} \int_0^{2\pi} n_e(x,t)\cos(mx) \, dx =
\frac{1}{\pi} \int_0^{2\pi}\cos\Bigl\{ m \Bigl[s + E(s,t) \Bigr] \Bigr\}
\, ds,\label{Bm1ft}
\end{equation}
and similarly,
\begin{equation}
B_{2m}(t) = \frac{1}{\pi} \int_0^{2\pi}\sin\Bigl\{ m \Bigl[s + E(s,t) \Bigr] \Bigr\}
\, ds,\label{Bm2ft}
\end{equation}
with $E(s,t)$ given by Eq.~(\ref{barE}).

Equation (\ref{nefx}) along with the first equation in (\ref{dimleqs}) integrated over
$dx$ leads to the Fourier expansion of $E(x,t)$:
\begin{equation}
\label{Efx}
E(x,t) = \sum_{m=1}^{\infty} \frac{1}{m} \Bigl[ - B_{1m}(t)\sin(m x) + B_{2m}(t)
\cos(m x) \Bigr].
\end{equation}
And finally, Eqs.~(\ref{dEdt}) and (\ref{dimlpar}) lead to
\begin{equation}
\label{neve}
n_e(x,t) v_e(x,t) =
\frac{\partial E(x,t)}{\partial t}.
\end{equation}
This equation along with (\ref{Efx}) and (\ref{nefx}) defines $v_e(x,t)$.

The freedom in the choice of the constants $A_{1n}$, $\varphi_{1n}$,  $ A_{2n}$, and
$\varphi_{2n}$ can be reduced from four to two if we prescribe the initial
electron density profile $n_{e0}(x) \equiv n_e(x, t=0)$. Putting $t = 0$ in
Eq.~(\ref{barne}) and using standard formulas for Fourier coefficients we
arrive at ($n_e^{-1} \, ds = dx$)
\begin{eqnarray}
A_{1n} g_{1n}(0) &=& -\frac{1}{n\pi} \int_0^{2\pi}% \!\!
\cos\Bigl\{ n \bigl[ x - E(x,0) \bigr] \Bigr\} \, dx\nonumber\\
&\equiv& -\frac{1}{n\pi} \int_0^{2\pi} \!\!
\cos\Bigl[ n \int_0^x n_{e0}(x')\,dx' \Bigr] \,
dx,\label{AnBm}
\end{eqnarray}
where $E(x,0)$ is given by Eq.~(\ref{Efx}) and we have chosen $s(0,0) = 0$.
An analogous formula for $A_{2n} g_{2n}(0)$ can be obtained from Eq.~(\ref{AnBm})
if we drop the minus signs in front of the integrals $dx$ and replace the cosine
functions by sines.

In general the integrals in Eqs.~(\ref{Bm1ft}), (\ref{Bm2ft}) and (\ref{AnBm})
(first line) are not expressible in terms of elementary functions, but if
the sum over $m$ or $n$ is truncated at some $M$ or $N$, one can
calculate as many integrals as needed (numerically).

If either $M=1$ or $N=1$ (first harmonics only), and additionally $B_{21}(0) = 0$
or $A_{21} = 0$ (i.e., $n_{e0}(x)$ or $n_e^{-1}(s,t)$ as a function of $s$ are
even functions) the relevant coefficients are expressible in terms of Bessel
functions. Thus using the identity \cite{rizik} (p. 423)
\begin{equation}
\int_0^{\pi} \cos(ax - z\sin x)\,dx = \pi J_a(z),\label{ident}
\end{equation}
we obtain
\begin{equation}
A_{1n} g_{1n}(0) = \frac{2}{n} (-1)^{n+1}
J_n\bigl(nB_{11}(0)\bigr),\label{Anp}
\end{equation}
and $A_{2n} \equiv 0$, if $B_{1m}(0) = 0$ for $m > 1$ and $B_{2m} \equiv 0$.

Similarly,
\begin{equation}
B_{1m}(t) = 2 J_m\bigl[ m A_{11} e^{-\alpha_1 t} g_{11}(t)\bigr],\label{Bmp}
\end{equation}
and $B_{2m}(t) \equiv 0$, if $A_{1n} = 0$ for $n > 1$ and $A_{2n} \equiv 0$.

\section{\label{applc}Applicability conditions}

As for applicability of Eqs.(\ref{nue}), (\ref{eta}), one assumes that
the plasma is quasineutral ($n_e \approx n_i$) and
the distribution functions are not far from local Maxwellians. Other
approximations come from standard assumptions of the Chapman--Enskog method.
Therefore, uncertainty factors of two or three cannot be excluded \cite{brag}.

If the requirement that the electron pressure term in Eq.~(\ref{mom}) is
negligible as compared to the electric force is expressed in terms of the
dimensionless quantities (\ref{dimlpar}), it takes the form
\begin{equation*}
3 n_e k \, \lvert\frac{\partial n_e}{\partial x}\rvert T_e  \ll 
\frac{m_e\omega_{pe}^2}{k} \lvert E \rvert, \quad \text{where} \quad
\frac{\partial n_e}{\partial x} =
n_e \frac{\partial n_e}{\partial s},
\end{equation*}
in view of (\ref{dimleqs}). Differentiating Eq.~(\ref{barne})
$\partial/\partial s$ and comparing the result with Eq.~(\ref{barE}) it
can be seen that 
\begin{equation*}
n_e^{-2}(s,t) \frac{\partial n_e}{\partial s}  \approx E(s,t),
\end{equation*}
where both sides differ in higher space harmonics only. Finally the requirement in
question takes the form
\begin{equation}
n_e^4 \Bigl(\frac{3\pi}{e^2}\Bigr) \frac{T_e}{n_0 \lambda^2}  \ll 1,
\label{valc2}
\end{equation}
which is the non-linear counterpart of condition (\ref{valc1}). In the linearised
theory, where the dimensionless $n_e \approx 1$, this condition takes the form
of Eq.~(\ref{valc1}).
Expressing $T_e$ in eV and calculating $T_e[\text{eV}]/(n_0 \lambda^2)$ from
Eq.~(\ref{nub}), we end up with
\begin{equation}
n_{e\,\text{max}}^4 \, \nu\,\eta \ll 0.04.\label{Tmax}
\end{equation}
This indicates that at least one of the coefficients $\nu$ or $\eta$ must be
much smaller than $0.2$ ($n_{e\,\text{max}} > 1$). Note that for given plasma
parameters $n_0$ and $T_e$,
$\eta$ is defined by Eq.~(\ref{etab}) and $\nu$ is related to $\lambda$ by
Eq.~(\ref{nub}), i.e., either $\nu$ or $\lambda$ can be prescribed.

\begin{table*}[t!]
\caption{\label{tableta}Dimensionless plasma resistivity $\eta$ and square root of
dimensionless viscosity times the wavelength $\sqrt{\nu}\,\lambda$ in some typical plasmas.
Choosing some value of $\nu\ (<1)$ and multiplying the last column by $\sqrt{\nu^{-1}}$ one
obtains $\lambda$ in cm.}
\begin{ruledtabular}
\begin{tabular}{cccclc}
Plasma Type  & $n_0$ cm$^{-3}$ & $T_e$ eV & $\lambda_D$ cm & \hspace{2em}$\eta$ &
$\lambda_{\text{min}}\equiv \sqrt{\nu}\,\lambda$ cm\\[1ex] \hline
Interstellar plasma & $1$       & $1$    & $7 \times 10^2$   & $3.08 \times 10^{-10}$ & $9.37 \times 10^7$\\
Gaseous nebula      & $10^3$    & $1$    & $20$              & $8.30 \times 10^{-9}$  & $5.71 \times 10^5$\\
Solar corona        & $10^6$    & $10^2$ & $7$               & $3.03 \times 10^{-10}$ & $9.44 \times 10^5$\\
Diffuse hot plasma  & $10^{12}$ & $10^2$ & $7 \times 10^{-3}$& $2.12 \times 10^{-7}$  & $35.7$\\
Solar atmosphere, gas discharge & $10^{14}$ & $1$    & $7 \times 10^{-5}$ & $9.59 \times 10^{-4}$
& $5.31 \times 10^{-3}$\\
Warm plasma      & $10^{14}$ & $10$   & $2 \times 10^{-4}$ & $4.47 \times 10^{-5}$ & $7.78 \times 10^{-2}$\\
Hot plasma       & $10^{14}$ & $10^2$ & $7 \times 10^{-4}$ & $1.82 \times 10^{-6}$  & $1.22$\\
Thermonuclear plasma& $10^{15}$ & $10^4$ & $2 \times 10^{-3}$ & $7.20 \times 10^{-9}$  & $61.3$\\
Theta pinch      & $10^{16}$ & $10^2$ & $7 \times 10^{-5}$ & $1.52 \times 10^{-5}$  & $4.22 \times 10^{-2}$\\
Dense hot plasma & $10^{18}$ & $10^2$ & $7 \times 10^{-6}$ & $1.22 \times 10^{-4}$  & $1.49 \times 10^{-3}$\\
Laser plasma     & $10^{20}$ & $10^2$ & $7 \times 10^{-7}$ & $9.13 \times 10^{-4}$  & $5.44 \times 10^{-5}$
\end{tabular}
\end{ruledtabular}
\end{table*}

In Table~\ref{tableta} we present $\eta$ and $\sqrt{\nu}\,\lambda$ as functions of $n_0$
and $T_e$. As oscillating results
are obtained only for $\nu<1$, values of the wavelength $\lambda$ must be greater than
$\lambda_{\text{min}} \equiv \lambda(\nu = 1)$ given in the last column of
Table~\ref{tableta}. Values of $\lambda$ in cm can be
obtained by multiplying the last column of Table~\ref{tableta} by $\sqrt{\nu^{-1}}$. For
example, for $\nu = 10^{-10}$, $\sqrt{\nu^{-1}} = 10^5$ and the last column of Table~%
\ref{tableta} gives the values of $\lambda$ in km. Similarly for $\nu = 10^{-4}$, it
gives $\lambda$ in m, etc.

Table \ref{tableta} indicates that for realistic plasmas, $\eta$ is always much smaller
than one. Therefore the requirement (\ref{Tmax}) is nearly always fulfilled.
Thus, if $\nu$ ($< 1$) is close to unity so that the viscous term in Eq.~(\ref{veql})
is significant, both quasi-neutrality and linearisation of Eq.~(\ref{veql}) require that
the dimensionless $n_e(x,t)$ cannot be significantly different from unity
($n_{e\,\text{max}}^4 \sim 1$).
In the opposite limit of $\nu \ll 1$, the viscous term in Eq.~(\ref{veql}) is
insignificant and there is no need to worry about its linearisation or quasi-neutrality.
In that case $n_{e\,\text{max}}$ can be large, but condition Eq.~(\ref{Tmax}) will be
fulfilled for reasonable values of $n_{e\,\text{max}}$ (if $n_{e\,\text{max}}^4 \, \nu <
1$).

If in a realistic plasma (where $\eta$ is always small) we also choose $\nu \ll 1$,
the plasma behaviour will be well approximated by an idealized model of a plasma being strictly
cold and collisionless ($T_e = \nu = \eta = 0$). In that case, Eqs.~(\ref{dimles}) with
$\alpha_n \equiv 0$ and $\omega_n \equiv 1$ give an \textit{exact}
and \textit{general} (periodic) solution of our non-linear initial equations
(\ref{cont})--(\ref{pois}):
\begin{subequations}
\label{dimles0}
\begin{eqnarray}
x(s,t) &=& s + E ,\label{barx0}\\
E(s,t) &=& \sum_{n=1}^{\infty} \: \bigl[ A_{1n}\cos(t + \varphi_{1n})\sin(ns) 
+ A_{2n}\cos(t + \varphi_{2n})\cos(ns)\bigr],\label{barE0}\\
n_e^{-1}(s,t) &=& 1 + \sum_{n=1}^{\infty} \: n
\bigl[ A_{1n}\cos(t + \varphi_{1n})\cos(ns)
- A_{2n}\cos(t + \varphi_{2n})\sin(ns)\bigr],\label{barne0}\\
v_e(s,t) &=& - \sum_{n=1}^{\infty} \: 
\bigl[ A_{1n}\sin(t + \varphi_{1n})\sin(ns) 
+ A_{2n}\sin(t + \varphi_{2n})\cos(ns)\bigr].\label{vsumd0}
\end{eqnarray}
\end{subequations} 

As $E(s,t)$ in Eqs.~(\ref{barx0}), (\ref{barE0}) is a periodic function of $t$, the
Fourier coefficients $B_{1m}(t)$ and $B_{2m}(t)$ given by (\ref{Bm1ft}) and (\ref{Bm2ft})
are also periodic, with period $\pi$. This in turn implies the same periodicity of all
plasma parameters. They will all oscillate, with frequency (in real time) equal
to the plasma frequency $\omega_{pe}$. The fact that this frequency is independent of the
amplitude of nonlinear oscillations is a characteristic feature of an ideal plasma, i.e.,
of any real plasma for which $\nu,\eta \ll 1$. This fact is not new but is demonstrated
here for the general solution (\ref{dimles0}).

Another characteristic feature of the general solution (\ref{dimles0}) is that the
dimensionless $n_e(x,t)$ is always greater than $\tfrac{1}{2}$ and blows up if $n_e(x,t)
\to \tfrac{1}{2}$ somewhere. One obtains $n_e \to \infty$ when the sum in
Eq.~(\ref{barne0}) tends to $-1$ (its unreachable minimum) while its value at time shifted
by $\pi$ (the half period) tends to $1$ (its unreachable maximum). The fact that in a cold
and collisionless plasma $n_{e\,\text{min}}$ must be greater than $\tfrac{1}{2}$ must be
kept in mind when prescribing the initial electron density profile $n_{e0}(x) \equiv n_e(x,
t=0)$. Any minimum in the initial density distribution tending to $\tfrac{1}{2}$ produces an
exploding maximum after a half period of plasma oscillation. However, when the plasma
viscosity $\nu$ is increased so as to introduce a noticeable damping, growth of the maximum
in question is reduced, and the corresponding minimum in the initial distribution
\textit{can} fall below $\tfrac{1}{2}$. This will be illustrated in Section VII.

Table \ref{tableta} indicates that for a typical thermonuclear plasma with $T_e = 10$ keV,
$\eta$ is much smaller than one and so condition (\ref{Tmax}) for the plasma being
``cold'' is fulfilled if $n_{e\,\text{max}}^4 \, \nu < 1$. This would be so even if one could
increase the number density $n_0$ of such a plasma to the value reached in a laser plasma,
$n_0 = 10^{20}$ cm$^{-3}$, which would result in $\eta = 1.52 \times 10^{-6}$.


\section{Examples}

In view of the fact that for realistic plasmas $\eta$ is always much smaller than
one, the actual value of $\eta$ will have no visible effect on the plots of $n_e(x,t)$,
etc. where it will be taken equal to zero. The plots will depend on whether or not
$\nu$ is much smaller than one. In any case, values of $\lambda$ in cm can be
obtained by multiplying the last column of Table~\ref{tableta} by $\sqrt{\nu^{-1}}$.

In Figs.~\ref{A1neq0} and \ref{A1A2neq0} we present typical examples of the electron
density evolution for a small number of modes included in Eqs.~(\ref{dimles}). Various
spatially periodic structures can be produced but the initial density distribution
follows.

\begin{figure}[t!]
\begin{center}

\ifpdf
    \includegraphics[type=png,ext=.png,read=.png,scale=0.6]{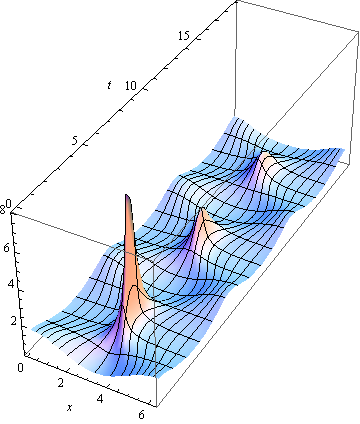}
\else
    \includegraphics[type=eps,ext=.eps,read=.eps,scale=0.9]{B1a}
\fi 
%\centerline{\includegraphics[scale=.39]{B1a.eps}}
\caption{(color online). Plot of $n_e(x,t)$ when only
$B_1(0) \equiv B_{11}(0) = 0.55$ is nonzero, $\nu = 0.015$ and $N=40$.
($\sqrt{\nu^{-1}} = 8.16$, see Table~\ref{tableta}.)
Note that in the presence of even weak viscosity $B_1(0)$ can exceed $1/2$,
see Fig.~\ref{minnu}.\label{B1smv}
}
\end{center}
\end{figure}

\begin{figure}[h!]
\begin{center}

\ifpdf
    \includegraphics[type=png,ext=.png,read=.png,scale=0.6]{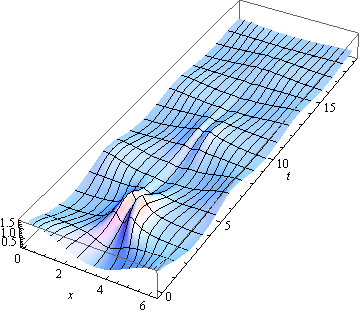}
\else
    \includegraphics[type=eps,ext=.eps,read=.eps,scale=0.9]{B1b}
\fi 
%\centerline{\includegraphics[scale=.35]{B1b.eps}}
\caption{(color online). Plot of $n_e(x,t)$ when only
$B_1(0) \equiv B_{11}(0) = 0.55$ is nonzero, $\nu = 0.1$ and $N=40$.
($\sqrt{\nu^{-1}} = 3.16$, see Table~\ref{tableta}.)
The observed bifurcation of the maximum under strong viscosity is a new
nonlinear effect.\label{B1lgv}
}
\end{center}
\end{figure}

\begin{figure}[h]
\begin{center}

\ifpdf
    \includegraphics[type=png,ext=.png,read=.png,scale=0.6]{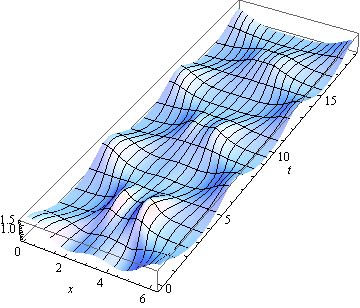}
\else
    \includegraphics[type=eps,ext=.eps,read=.eps,scale=0.9]{B1B2}
\fi 
%\centerline{\includegraphics[scale=.9]{B1B2.eps}}
\caption{(color online). Plot of $n_e(x,t)$ when only
$B_{11}(0) = 0.4$ and $B_{12}(0) = 0.3$ are nonzero, $\nu = 0.015$ and $N=40$.
($\sqrt{\nu^{-1}} = 8.16$, see Table~\ref{tableta}.)\label{B1B2lgv}
}
\end{center}
\end{figure}

We can prescribe the initial density distribution $n_e(x, t=0)$ either directly, or by
defining initial values of its Fourier coefficients $B_{1m}(0)$, $B_{2m}(0)$. In the
latter case periodicity is guaranteed and in any case, the Fourier coefficients
$A_{1n}$ and $A_{2n}$ can be calculated by using Eq.~(\ref{AnBm}). In the simplest case
of only the first harmonic present, $A_{1n}$ are expressible in terms of Bessel functions,
see Eq.~(\ref{Anp}). Otherwise numerical integration is necessary which, however, is
a simple numerical task. Examples are shown in Figs.~\ref{B1smv}--\ref{B1B2lgv}.

The particular solution (\ref{dimles}) in which $A_{1n}$ is given by (\ref{Anp}) with
$\varphi_{1n} = -\pi/2$,
reduces to that of \cite{david} if $\nu = \eta = 0$, though in a different
notation. Denoting, as in \cite{prl09}, the amplitude $B_{11}(0)$ of the initial density
deviation from equilibrium by $B_1(0)$, the known condition $B_1(0) < \tfrac{1}{2}$ follows
from the identity $\sum_{n=1}^{\infty} J_n(n z) = \frac{z}{2(1 - z)}$
\cite{rizik} (p. 935). For $B_1(0) = \tfrac{1}{2}$, $n_e(x, t=0)$ reaches its
minimum equal $\tfrac{1}{2}$ at $x=\pi$ and $n_e^{-1}$ becomes zero at $s = t = \pi$
($n_e \to \infty$). This illustrates the general property of an ideal plasma formulated
at the end of Sec.~VI.

\begin{figure}[t!]
\begin{center}

\ifpdf
    \includegraphics[type=png,ext=.png,read=.png,scale=0.28]{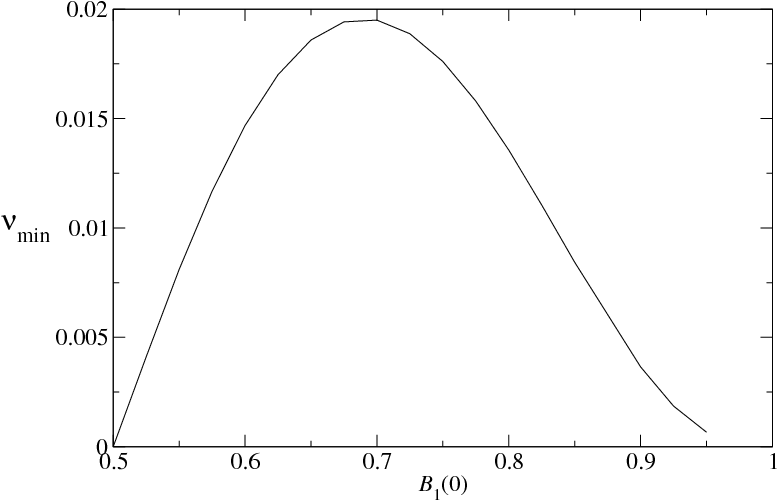}
\else
    \includegraphics[type=eps,ext=.eps,read=.eps,scale=0.28]{bmax1}
\fi 
%\centerline{\includegraphics[scale=.28]{bmax1.eps}}
\caption{ Minimal admissible value of $\nu$  as a function of $B_1(0)$.\label{minnu}
}
\end{center}
\end{figure}

The behavior of the solution in question for $\nu > 0$ is shown in
Figs.~\ref{B1smv} and \ref{B1lgv}. In a viscous plasma $B_1(0)$ can exceed $1/2$,
see Fig.~\ref{minnu}. This figure assumes $\eta \ll 1$ and gives the minimal
admissible value of $\nu$ for given $B_1(0)$. For $\nu =
\nu_{\text{min}}\bigl[B_1(0)\bigr]$, the smallest minimum of $n_e^{-1}
(x=\pi, t)$ at $t = t_{\text{min}}$ becomes zero.
Note also that if $\nu$ is sufficiently
large, see Fig.~\ref{B1lgv}, where $\nu = 0.1$, a new nonlinear
effect can be noticed, i.e., the largest density maximum splits in two, with a saddle
point between the peaks. This effect is due to the presence of the integer $n^2$ in
front of $\nu$ in Eq.~(\ref{alphab}).

\begin{figure}[b!]
\begin{center}

\ifpdf
    \includegraphics[type=png,ext=.png,read=.png,scale=0.65]{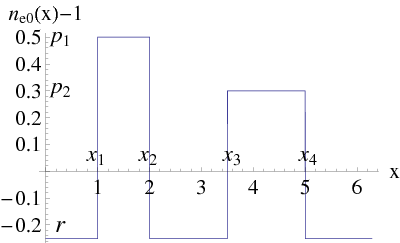}
\else
    \includegraphics[type=eps,ext=.eps,read=.eps,scale=0.55]{stepf}
\fi 
%\centerline{\includegraphics[scale=.55]{stepf.eps}}
\caption{An example of possible initial electron density deviation from equilibrium,
$n_e(x)-1$. Here: $x_1=1$,
$x_2=2$, $x_3=3.5$, $x_4=5$, $p_1=0.5$, $p_2=0.3$, and $r=-0.251$ as calculated
from (\ref{constr}).\label{stepf}
}
\end{center}
\end{figure}
\begin{figure}[t!]
\begin{center}

\ifpdf
    \includegraphics[type=png,ext=.png,read=.png,scale=0.75]{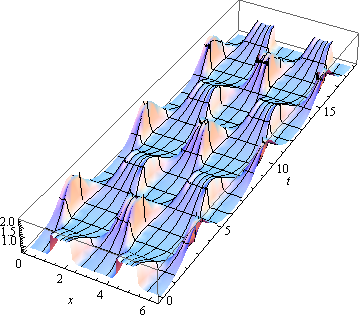}
\else
    \includegraphics[type=eps,ext=.eps,read=.eps,scale=0.9]{stfpp}
\fi 
%\centerline{\includegraphics[scale=.9]{stfpp.eps}}
\caption{(color online). Plot of $n_e(x,t)$ for a two pulse
profile, $p_1=p2=0.7$, $\nu = 0$ and $N=40$. See caption to Fig.~\ref{A1neq0}
for the consequences of $\nu=0$.\label{stfpp}
}
\end{center}
\end{figure}
\begin{figure}[b!]
\begin{center}

\ifpdf
    \includegraphics[type=png,ext=.png,read=.png,scale=0.75]{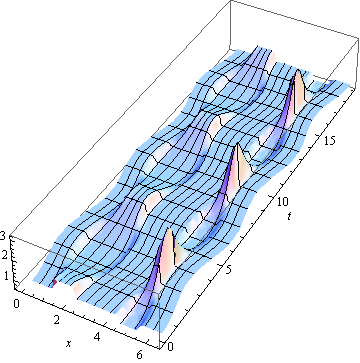}
\else
    \includegraphics[type=eps,ext=.eps,read=.eps,scale=0.9]{stfpn}
\fi 
%\centerline{\includegraphics[scale=.9]{stfpn.eps}}
\caption{(color online). Plot of $n_e(x,t)$ for a two pulse 
profile, $p_1=0.5$, $p2=-0.4$, $\nu = 0$ and $N=100$.\label{stfpn} 
}
\end{center}
\end{figure}
\begin{figure}[t!]
\begin{center}

\ifpdf
    \includegraphics[type=png,ext=.png,read=.png,scale=0.8]{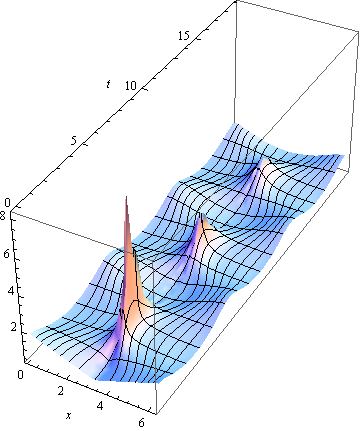}
\else
    \includegraphics[type=eps,ext=.eps,read=.eps,scale=0.9]{sawsn}
\fi 
%\centerline{\includegraphics[scale=.9]{sawsn.eps}}
\caption{(color online). Plot of $n_e(x,t)$ for a saw-tooth initial density
profile, $b=0.63$, $\nu = 0.015$ and $N=40$. ($\sqrt{\nu^{-1}} = 8.16$,
see Table~\ref{tableta}.)
Note that in the presence of even weak viscosity $b$ can exceed $\tfrac{1}{2}$,
i.e., the minimum of $n_e(x, t=0)$ can fall below $\tfrac{1}{2}$.\label{sawsn}
}
\end{center}
\end{figure}
\begin{figure}[b!]
\begin{center}

\ifpdf
    \includegraphics[type=png,ext=.png,read=.png,scale=0.8]{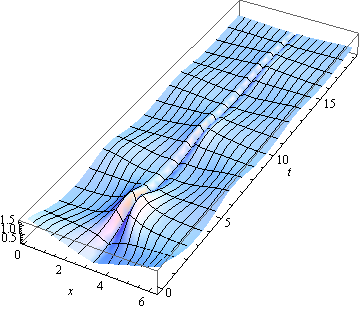}
\else
    \includegraphics[type=eps,ext=.eps,read=.eps,scale=0.9]{sawln}
\fi 
%\centerline{\includegraphics[scale=.9]{sawln.eps}}
\caption{(color online). Plot of $n_e(x,t)$ for a saw-tooth initial density
profile, $b=0.55$, $\nu = 0.1$ and $N=40$. ($\sqrt{\nu^{-1}} = 3.16$,
see Table~\ref{tableta}.)
Again the  bifurcation of the maximum under strong viscosity can be seen.\label{sawln}
}
\end{center}
\end{figure}

The integral $dx$ in Eq.~(\ref{AnBm}) is elementary if $n_{e0}(x)$
is a sequence of step functions. An example of this is shown in Fig.~\ref{stepf},
where $n_{e0}(x) - 1$ is given in the form of two pulses, $p_1$ and $p_2$,
emerging from a reference level $r$. The parameters $p_1$, $p_2$, and $r$ can be
positive, negative or zero but none of them can be smaller than $-1$, as that would
make $n_{e\,0} < 0$. Using Eq.~(\ref{AnBm}) (the second line) we easily find:
\begin{eqnarray}
A_{1n} g_{1n}(0) &=& -\frac{1}{\pi n^2} \Bigl\{
\frac{1}{1 + r} \bigl[ \sin(n s_1) + \sin(n s_3) - \sin(n s_2) - \sin(n s_4) \bigr]\nonumber\\
&&+ \frac{1}{1 + p_1}
\bigl[ \sin(n s_2) - \sin(n s_1) \bigr] + \frac{1}{1 + p_2} \bigl[ \sin(n s_4) -
\sin(n s_3) \bigr] \Bigr\},\label{A1nco}
\end{eqnarray}
where
\begin{eqnarray*}
s_1 &=& (1 + r) x_1,\\
s_2 &=& s_1 + (1 + p_1)(x_2 - x_1),\\
s_3 &=& s_2 + (1 + r)  (x_3 - x_2),\\
s_4 &=& s_3 + (1 + p_2)(x_4 - x_3).
\end{eqnarray*}
$A_{2n} g_{2n}(0)$ will be given by the right hand side of
Eq.~(\ref{A1nco}) if we replace $\sin$ by $\cos$.

The integral $\int_0^{2\pi}(n_{e0} - 1) \, dx$ must be zero, leading to a
constraint on the independent parameters. Solving it for $r$ we obtain
\begin{equation}
r = -\frac{p_1 (x_2 - x_1) + p_2 (x_4 - x_3)}{2\pi - (x_2 - x_1) - (x_4 - x_3)}.
\label{constr}
\end{equation}

Examples are shown in Figs.~\ref{stfpp} and \ref{stfpn}.

If the integral $dx'$ in Eq.~(\ref{AnBm}) is
expressible analytically, the integral $dx$ in this equation  can easily
be found numerically. For example, for the saw-tooth initial density profile
we obtain ($\varphi_{1n} = -\pi/2$):
\begin{equation}
A_{1n} = - \frac{2}{n\pi} \int_0^{\pi} \cos\Bigl\{n x\bigl[(1 + b -
b x/\pi)\bigr]\Bigr\} \, dx, \quad A_{2n} \equiv 0,\label{sawt}
\end{equation}
where $b$ is the ``amplitude'' of the initial density perturbation
($n_{e\,0}(x=0) = 1 + b$). The results are shown in Figs.~\ref{sawsn} and
\ref{sawln}.

Note that if we linearise Eq.~(\ref{veq}), by neglecting $v_e\partial v_e/\partial x$ on
the left hand side and replacing $n_e$ by $n_0$ on the right hand side, the resulting
equation will be identical with the linearised form of Eq.~(\ref{veql}) if we replace
$s$ by $x$ and $\tau$ by $t$. This means that
$v_e(x,t)$ for the linearised problem will be given by Eq.~(\ref{vsum}), or its
dimensionless equivalent Eq.~(\ref{vsumd}), if we replace $s$ by $x$ and $\tau$ by $t$.
Then, using the third equation in (\ref{dimleqs}) (with $\tau \to t$) and the first
equation we can find $E(x,t)$ and $n_e(x,t)$. Finally, $E(x,t)$ will be given by
Eq.~(\ref{barE}) with $s$ replaced by $x$, and for $n_e$ we will get
\begin{equation}
n_e(x,t) = 1 + \sum_{n=1}^{\infty}
e^{-\alpha_{n} t}n\bigl[ A_{1n} g_{1n}(t) \cos(nx)
- A_{2n} g_{2n}(t) \sin(nx)\bigr].\label{barnex}
\end{equation}
Putting here and in Eqs.~(\ref{barE}) and (\ref{vsumd}) $A_{1n} = 0$ for $n > 1$,
$A_{2n} \equiv 0$, and $\varphi_{11} = - \pi/2$ we obtain
\begin{subequations}
\begin{eqnarray}
E(x,t) &=& - A_{11} G_1(t) \sin x,\label{Elin}\\
n_e(x,t) &=& 1 + A_{11} G_1(t) \cos x,\label{nelin}\\
v_e(x,t) &=& A_{11} e^{-\alpha_1 t} \,
\frac{\sin(\omega_1t)}{\omega_1} \sin x,\label{velin}\\
G_1(t) &=& e^{-\alpha_1 t} \Bigl[\alpha_1
\frac{\sin(\omega_1t)}{\omega_1} + \cos(\omega_1t)\Bigr],
\end{eqnarray}
\end{subequations}
where $\omega_1 = \sqrt{1 - \alpha_1^2}$, $\alpha_1 = \nu + \eta$.
This result becomes a linear counterpart of the nonlinear solution discussed
earlier (only $B_{11}(0)$ nonzero) if we put $A_{11} = B_1(0)$. The amplitude
$A_{11}$ can take any value between $-1$ and $1$, but there is no maximum
amplification shown in Fig.~\ref{B1smv}. There is also no maximum splitting
for large enough viscosity shown in Fig.~\ref{B1lgv}. Both are clearly nonlinear
effects, not present in the linearised theory.

\section{Conclusions and final remarks}

The 1D Langmuir waves are importantly recently revisited in connection with
laser-driven plasma-based electron accelerators. A full non-linear fluid
description of these waves, with the dissipative effects included, is given by
Eqs.~(\ref{dimles}). They  define the dynamics of the macroscopic parameters of
the wave in a simple parametric form, only containing trigonometric and exponential
functions.
This simple form can be used to represent graphically the time evolution of the wave,
$n_e(x,t)$, etc., while direct analytical formulas for these quantities,
Eqs.~(\ref{nefx})--(\ref{neve}), are much more complicated.

Our analysis indicates that in real plasmas, both space and laboratory, the
influence of plasma resistivity and electron pressure forces on the waves in question
is negligible.

The role of electron viscosity can be noticeable as one decreases the wavelength
$\lambda$ of the Langmuir wave excited in a plasma. However, for $\lambda$
sufficiently large as compared to $\lambda_{\text{min}}$ given in TABLE I (that is
the dimensionless viscosity coefficient $\mu$ sufficiently small compared to unity),
the effect of viscosity is also negligible. In that case, the
waves are well described by an idealized model of a plasma being strictly cold and
collisionless ($T_e = \nu = \eta = 0$). Within this model,  all plasma parameters
oscillate in time with the plasma frequency $\omega_{pe}$ (\ref{omegape}),
independently of the initial density shape or its amplitude.
Furthermore, the electron number density $n_e$ is always
greater than half its equilibrium value $\tfrac{1}{2} n_0$. Any minimum in the
initial electron density distribution $n_e(x,t=0)$ tending to $\tfrac{1}{2} n_0$
would produce an exploding maximum after half a period of the plasma oscillation.
These facts were known for a purely cosinusoidal deviation of the initial electron
number density from its equilibrium value $n_0$, but are demonstrated here for
arbitrary deviation.

When increasing the wavelength $\lambda$ (i.e., for $\nu$ small but finite, e.g.,
$\nu = 0.015$), growth of the maximum is reduced, and the minimum can
fall below $\tfrac{1}{2} n_0$, see Figs.~\ref{B1smv} and \ref{sawsn}. And for $\nu$
relatively large, e.g., $\nu = 0.1$, the maximum bifurcates as shown in
Figs.~\ref{B1lgv} and \ref{sawln}.

Our requirement that $n_e/n_0$ should not be considerably larger than unity
concerns the accuracy of the viscous force (for validity of linearisation and
quasi-neutrality). However, this term only corrects the motion, while the driving
electric force and the convective non-linearities are taken into account exactly.
In fact, only for relatively large $\nu$, must one stipulate that $n_e$ should not
exceed $n_0$ too much.

\end{document}